\documentclass[manuscript)]{aastex}

\usepackage{lscape,graphicx,epsfig,natbib,color}
\usepackage{epstopdf}
\makeatletter

\newcommand{\Rmnum}[1]{\expandafter\@slowromancap\romannumeral #1@}
\makeatother

\slugcomment{Submitted to ApJ}

\shorttitle{High resolution He\,\textsc{I} 10830 \AA\ narrow-band imaging of an M-class flare. I -
analysis of sunspot dynamics during flaring}
\shortauthors{Wang et al.}

\begin{document}
\title{High resolution He\,\textsc{I} 10830 \AA\ narrow-band imaging of an M-class flare. I -
analysis of sunspot dynamics during flaring}

\author{Ya Wang\altaffilmark{1,2}, Yingna Su\altaffilmark{1}, Zhenxiang Hong\altaffilmark{1,2}, Zhicheng Zeng\altaffilmark{3}, Kaifan Ji\altaffilmark{4}, Philip, R. Goode\altaffilmark{3}, Wenda Cao\altaffilmark{3} and Haisheng Ji\altaffilmark{1,3} }

\affil{$^1$ Key Laboratory of DMSA, Purple Mountain Observatory, CAS, Nanjing, 210008, China}
\affil{$^2$ University of CAS, Beijing 100049, China}
\affil{$^3$ Big Bear Solar Observatory, 40386 North Shore Lane, Big Bear City, CA 92314, USA}
\affil{$^4$ Yunnan Astronomical Observatories, Kunming 650011, China}

\begin{abstract}
In this paper, we report our first-step results of high resolution He\,\textsc{i} 10830 \AA\ narrow-band imaging (bandpass: 0.5 {\AA}) of an M1.8 class two-ribbon flare on July 5, 2012. The flare was observed with the 1.6 meter aperture New Solar Telescope at Big Bear Solar Observatory. For this unique data set, sunspot dynamics during flaring were analyzed for the first time. By directly imaging the upper chromosphere, running penumbral waves are clearly seen as an outward extension of umbral flashes, both take the form of absorption in the 10830 \AA\ narrow-band images. From a space-time image made of a slit cutting across a flare ribbon and the sunspot, we find that the dark lanes for umbral flashes and penumbral waves are obviously broadened after the flare. The most prominent feature is the sudden appearance of an oscillating absorption strip inside the ribbon when it sweeps into the sunspot's penumbral and umbral regions. During each oscillation, outwardly propagating umbral flashes and subsequent penumbral waves rush out into the inwardly sweeping ribbon, followed by a returning of the absorption strip with similar speed.  We tentatively explain the phenomena as the result of a sudden increase in the density of ortho-Helium atoms in the area of the sunspot being excited by the flare's EUV illumination. This explanation is based on the observation that 10830 \AA\  absorption around the sunspot area gets enhanced during the flare. Nevertheless, questions are still open and we need further well-devised observations to investigate the behavior of sunspot dynamics during flares.
\end{abstract}

\keywords{Sun: atmosphere --- Sun: line formation ---- Sun: sunspot --- Sun: flare}

\section{INTRODUCTION} \label{s-intro}
Helium is the second most abundant element in the Sun, as well as in the universe. Since its first excited energy levels are about 20 eV above the ground state, Helium lines are associated  with  extreme conditions like a high effective temperature or high energy photons in the solar atmosphere. From the ground, the only observable line among helium lines in the quiet solar disk is the near-infrared 10830 \AA\ line formed in the upper chromosphere (Andretta $\&$ Jones 1997), which appears as a weak absorption line in the continuum background. He\,\textsc{i} 10830{\AA} is formed by transitions between the metastable levels 1s2s ($^3$S) and the closely spaced 1s2p ($^3$P) triplet levels. Under normal conditions, two mechanisms can populate these metastable levels, they are photo-ionization-recombination mechanism (PRM) and collisional mechanism (CM). PRM assumes that UV and EUV radiation from the corona removes one electron from a Helium atom in the chromosphere (T $\le$ $10^4$ K), and then the electron cascades to excited levels, while direct collisional excitation from the ground state occurs in relatively hotter material (T $\ge$ $2 \times 10^4$ K). On the other hand, flare-produced non-thermal electrons can produce collisional ionization followed by recombination (CRM), which can be regarded as an important mechanism. Recently, CRM has received some observational support (e.g., Liu et al. 2013; Xu et al. 2016). For PRM, direct support comes from the fact that He\,\textsc{i} 10830{\AA} absorption depth increases with increasing coronal EUV irradiation over solar disk (e.g., Avrett et al. 1994; Andretta \& Jones 1997). Zeng et al. (2014) provided a detailed analysis of a C class flare observed with 10830 \AA\ narrow band (NB) imaging equipment on the 1.6 meter aperture New Solar Telescope at Big Bear Solar Observatory (NST/BBSO; Goode et al. 2010) and space data from the Atmosphere Imaging Assembly (AIA; Lemen et al. 2012) and the Extreme Ultraviolet Variability Experiment (EVE; Woods et al. 2012) instruments on-board the Solar Dynamics Observatory (SDO; Pesnell et al. 2012). From several approaches, it was determined that illumination by EUV photons is a dominant mechanism for the excitation of the Helium atoms that produced 10830 {\AA} emission. If the PRM mechanism really works, one expects that suddenly enhanced EUV photons during flares will not only produce 10830 \AA\ emission in the flaring region but will also enhance absorption in the surrounding quiet regions. However this phenomenon has never been observed so far.

Spectroheliograms or NB filtergrams 10830 \AA\ have been used routinely to explore the nature of the upper chromosphere, like the Chromospheric Helium Imaging Photometer (CHIP) (Elmore et al. 1998) at Mauna Loa Solar Observatory (MLSO), the Optical Solar Patrol Network (OSPAN) instrument (formerly Improved Solar Observing Optical Network (ISOON), see Neidig et al. 1998), Synoptic Optical Long-term Investigations of the Sun (SOLIS) (Harvey et al. 1996) at the National Solar Observatory, and Optical and the Near-infrared Solar Eruption Tracer (ONSET) of Nanjing University (Fang et al. 2013). With the advent of large aperture solar telescopes like NST and the addition of techniques of adaptive optics with the application of speckle reconstruction, NB imaging at He\,\textsc{i} 10830 {\AA} has proven to be an important tool for observing the interface region, which lies between the photosphere and corona (Ji, Cao and Goode 2012; Zeng, Cao and Ji 2013). The connection between EUV radiation and He\,\textsc{i} 10830 absorption makes the morphologies at the two wavelengths look very similar. Except for flaring times (Judge et al., 2015), the triplet is optically thin against solar disk, making the photosphere visible as background around solar disk center. Thus, a He\,\textsc{i} 10830 {\AA} NB filtergram consists of information about the photosphere, upper chromosphere and transition region. With He\,\textsc{i} 10830 NB imaging on the NST, Ji, Cao and Goode (2012) were able to trace small-scale activity from the photosphere to the transition region. They reported ultra-fine energy channels for heating the corona rooted in intergranular lanes. With the same set of data, Zeng, Cao and Ji (2013) studied a small surge event showing that it is produced by magnetic reconnection in an intergranular lane area powered by the rapid emergence of granule-scale magnetic field. With this experience, it is clear that He\,\textsc{i} 10830 {\AA} NB imaging is especially good for observing filaments and their eruption, since we could trace the fibrils of a filament directly down to the photosphere. Hopefully, we will be able to observe the role of the underlying photosphere in the formation and activation of filaments.

During its commissioning period, we carried out a series of He\,\textsc{i} 10830 NB imaging observations with NST, targeting active region's filaments. On 2012 July 5, we successfully observed a filament eruption accompanied by an M1.8 flare in active region NOAA 11515. We performed a detailed multi-wavelength analysis of this flare event. For this data set, the activation and onset of the filament eruption were observed with unprecedented high spatial resolution. The most prominent feature was the oscillating absorption strip inside a flare ribbon when it sweeps into the sunspot penumbral and umbral areas. This phenomenon was totally unexpected. In this paper, we report the first-step results concentrating on analyzing this feature.

In section 2, we give a detailed description of the data reduction, results and discussion are presented in sections 3 and 4, respectively, followed by a summary in section 5.

\section{OBSERVATION AND DATA ANALYSIS} \label{s-data}
The NST at BBSO has an off-axis design, which gives many advantages in high resolution observation by reducing stray light and offering an unobstructed pupil (Goode et al. 2010; Cao et al. 2010a). Long duration good seeing conditions at BBSO, combined with its high-order Adaptive Optics (AO) system helps deliver diffraction-limited images consecutive for several hours. This unique seeing feature makes NST a most suitable large-aperture telescope for the study of solar flares.

On 2012, July 5, the seeing conditions were basically good in the morning and became poor but still acceptable, with AO correction, in the afternoon. There were several active regions on solar disk.  NOAA 11515 was selected as the targeted region, even though it was approaching the limb. The primary reason for the decision was that this region was quite active and there was an active filament, which kept oscillating as seen from the movie made of full disk H$_\alpha$ images of the preceding hours. In addition, due to projection effects, near-limb observations for a filament eruption have a greater possibility of capturing the whole eruptive process and revealing more vertical structures in the filament. As expected, the oscillation turned into eruption and flaring at around 21:38:00 UT.

The observations were made with NB Lyot filters in He\,\textsc{i} 10830 {\AA} blue wing (-0.25 {\AA}, bandpass: 0.5 {\AA}) and in H$_\alpha$ 6563 {\AA} blue wing (0.65 {\AA}, bandpass: 0.25 {\AA}), and with a broad band filter (bandpass: 10 {\AA}) of TiO 7057 {\AA} lines. In this paper, we will solely analyze the 10830 \AA\ NB imaging data. The 10830 {\AA} Lyot filter was made by Nanjing Institute for Astronomical and Optical Technology. A high sensitivity HgCdTe CMOS infrared focal plane array camera (Cao et al. 2010b) was used to obtain the 10830 {\AA} data. Speckle bursts of 100 frames each were taken at 10 frames per second rate by the fast CCD-camera. A lucky (best frame) image was selected from each burst. The KISIP speckle reconstruction code (W{\"o}ger \& von der L{\"u}he 2007) was employed for image-reconstruction, and post-processing including a correction for de-stretching. The whole data set for 10830 \AA\ images started at 16:36:29 UT and lasted nearly 6 hours. In total, 2007 frames were obtained, with a cadence of about 10 seconds, pixel size of 0$\arcsec$.0875 with a field of view (FOV) of 90$\arcsec$$\times$90$\arcsec$.

In this paper, we will solely analyze the 10830 \AA\ NB imaging data around the flaring period (21:20:36-22:22:07 UT), for which, 356 frames were analyzed.  Frame selected images (lucky images: 343 frames, 21:00:01 - 21:59:50 UT) were also used in order to clarify some important phenomena. With sunspots as reference, we carried out alignment for all 10830 \AA\ images, the alignment procedure also included de-rotation of the field of view.

\section{RESULTS} \label{s-res}

For an overview of the flare, we made a movie of this event with well-aligned 10830 {\AA} NB images (see the online movie named as NST\_flare\_20120705.mp4). As anticipated, most of the eruptive processes were well-captured. Owing to the optically thin nature of the line, the photospheric continuum shines through. This makes photospheric features like sunspots, pores, and granules clearly visible in the images.  For the sunspot, running penumbral waves can be clearly seen in the movie. Another deeply impressive thing given by the movie is the very high resolution observation of the earlier dynamic behavior of the filament, and then the subsequent eruption and flaring process.  A closer look at the oscillating, with our unique 10830 {\AA} NB images, shows that many penumbra-rooted loop-like fibrils keep raising up, accompanied by correlated brightenings at their footpoints. The filaments were eventually erupted being accompanied with a two-ribbon flare.

Figure 1 presents selected snapshots of the 10830 \AA\ filtergrams around the maximum of the two-ribbon flare. The two ribbons undergo a separation motion, with one ribbon sweeping into the penumbral and umbral regions. For this ribbon, the most notable phenomenon is that an oscillating absorption strip appears inside it (for Figure 1, also please watch the on-line movie). To analyze the dynamic behavior of the absorption strip, we used a conventional space-time diagram, which is made by stacking, over time, a series of one-dimensional images obtained from integrating over the width of the slits. The slit is roughly 10 Mm in length and 0.26 Mm in width (corresponding to $164 \times 5 $ pixels$^2$), cutting vertically across the strip along the swaying direction of the absorption strip (for the position of the slice, see Figure 1). The space-time diagram made from speckle-reconstructed images is given in Figure 2, over which a GOES X-ray light curve with the soft X-ray emission in the 0.5-4 {\AA} channel being plotted for comparison purpose. It shows that the flare begins at about 21:38 UT and peaks two times at 21:42:15 UT and 21:45:47 UT, respectively. The times are marked with two consecutive arrows. Before flare onset, we can see a fishbone-like absorption pattern originated from the umbra and extending to the penumbral area as running penumbral waves. The fishbone-like absorption pattern in the umbral region is the signature of umbral flashes at 10830 {\AA}. We use dash-dotted straight line to delineate penumbral traveling waves (Figure 2), and by calculating the average slope of the lines, we determine that their phase speed is about 22.1 km s$^{-1}$.

After the flare occurs, the fishbone-like absorption in the space-time diagram turns into a series of sawtooth-like patterns, projected against the flare ribbon. The pattern  in the space-time diagram reflects the oscillation of the absorption strip in the movie. As the space-time diagram shows, the sawtooth pattern is clearly a continuation of the umbral flashes, but with part of material clearly going back. We analyzed the period of oscillation by using the method of wavelet analysis. The results of which are displayed in Figure 3. The fast components of the intensity obtained from two lines  along the space-time diagram (dotted lines in Figure 2) before and after the flare are shown in the first row, which are obtained by the total intensity after subtracting their slow components. The nearly constant period of the umbral and penumbral waves before the flare is about 3 minutes, which clearly corresponds to chromospheric oscillations above sunspot umbra. It is the same period for the sawtooth like pattern, though the 3-minute periodicity appears less remarkable after 22:00 UT.

By visually comparing the dark fishbone-like lanes, before and after the flare, in  the slit image, we find that the dark lanes are obviously broadened. The phenomenon was confirmed by a space-time diagram made of lucky images. At the same time, we find that the light curve at 10830 {\AA} for the sunspot area gets obviously dimmed after the flare. This means a sudden increase of overall absorption in that area. To show this, we choose five regions  (Figure 4) in the sunspot area to see the changes in brightness during flares. In order avoid to spurious effects of speckle-reconstruction, the light curves were obtained from lucky images rather than speckle-reconstructed images. The regions are marked with five color boxes in Figure 4. As it shows, the boxes include one umbral region, three different penumbral regions and one region covering the umbra and penumbra. All light curves are given in Figure 4, in which a red arrow points to the second peak time of the flare. They all begin to decrease after flare onset, reach their lowest level at the peak times of the flare, and then they slightly increase, but still are lower than the flux before the flare.  The drop in the average intensity is about 3.0-4.0\%.  As a further examination, a series of spatial profiles for the sunspot's brightness along a slice cutting through it before and after the flare are compared with its average profile before the flare. They are displayed in Figure 5, we can see that the sunspot's brightness profile, including its penumbra, is decreased after the flare. Due to numerous small-scale activities in the quiet region surrounding the sunspot, we cannot confirm flare-related absorption changes in these regions.

\section{DISCUSSION} \label{s-disc}

For the first time, we have carried out very high resolution 10830 {\AA} narrow band (NB) imaging of a filament eruption and successfully observed it.  Owing to long-duration good seeing conditions of the BBSO site, the targeted observations were able to be continuously sustained for nearly 6 hours. The whole data set enables us to follow the entire evolution of the active region, the filament as well as the flare - all with high spatial and temporal resolution.  We believe that this unique data set will improve our understanding on the activation, and even the initiation of filament eruption. A detailed analysis of the magnetic energy build-up and initiation processes will be given in a subsequent submission. As a first step, this paper reveals some unexpected by-products, which are presented as one of the novel findings of NST.

From the on-line movie, running penumbral waves can be seen running out from the sunspot. With conventional space-time diagram for a slit cutting through the sunspot umbra and penumbra, a series of fishbone-like absorption patterns can be seen starting from the umbral region and extending to the penumbra, as running penumbral waves. In this paper, the word ``absorption" means simply a decrease in the counts over the instrument's bandpass. The intensity can be affected by many factors, like doppler shift, width, line depth, and core emission. Remember that the observation was made at He I 10830 \AA\ blue wing (-0.25 \AA), the absorption in umbral flashes and penumbral waves actually reflects population enhancement and flowing out of neutral Helium atoms.

With no doubt, the absorption pattern in the umbra is the signature of umbra flashes. Umbral flashes are defined as repetitive brightenings of the umbral chromosphere in Ca\,\textsc{ii} H \& K and other lines with roughly three-minute periodicity since their discovery by Beckers \& Tallant (1969). The signature of umbra flashes in chromospheric line profile shows periodic blue and red shifts with large displacements, which contrasts with the strong emission peak in Ca\,\textsc{ii} H \& K lines core (Felipe et al. 2010). Usually, 10830 {\AA} observation samples the upper chromosphere. Limb observations of the quiet Sun indicate the existence of He I emission shell, which was measured as 1.1-2.1 Mm above the solar surface (Penn and Jones 1996, Ji, You and Fan, 1997 and references therein). The height may represent the mean penetration depth of EUV photons in the chromosphere. The existence of the running penumbral waves and umbral flashes at 1083 nm NBIs indicate that the enhanced EUV emission in and around active regions might cause a deeper penetration depth around sunspot area. The absorption pattern for umbral flashes and penumbral waves in the 10830 {\AA} blue wing NB imaging basically reflects the blue shift displacements in the oscillating sunspot's upper chromosphere. Spectroscopic observations show that umbral flashes and running penumbral waves are closely related oscillatory phenomena, i.e., running penumbral waves are extensions of umbral flashes (e.g. Rouppe van der voort et al. 2003). With direct high resolution NB imaging in 10830 {\AA}, our observations provides a much clearer evidence for the continuous transition from umbral dark flashes to penumbral waves.

After the flare, the fishbone-like absorption strip in the space-time diagram becomes broadened, this is an interesting phenomenon, which immediately demands an explanation. We accept that umbral flashes and penumbral wave structures are extended upward propagating magneto-acoustic shock waves (Lites 1986; Rouppe van der voort et al. 2003; Yurchyshin et al. 2015; Tian et al. 2014). The broadened absorption strip reflects the broadened width of shock waves' front. Meanwhile, we find that 10830 {\AA} absorption in the sunspot area gets enhanced during the flare, as seen from temporal and spatial profile for the intensity of the sunspot. The two phenomena are certainly related and flare-associated. To our thoughts, the enhanced absorption can be attributed to the EUV illumination. To support the hypothesis, we extend the time period to include two micro-flares (precursors), which peaked at 21:20:30 and 21:31:00 UT, respectively.  The supporting results are given in Figure 6. From the on-line movie, we find that there occurred many small spatial scale pre-flare brightenings around the site of blue boxes in panel S1. The time profile for the small spatial scale 10830 \AA\ brightenings is given with a blue color light curve in panel L1. Among these brightenings, two are associated with prominent EUV irradiance around the same place. A sample of the two corresponding EUV peaks is given in panel L2 for AIA 304 \AA\ images. For other AIA EUV wavelengths, they have very similar time profiles.  The two EUV peaks have their corresponding signatures in GOES light curves (Figure 2), so they can be regarded as two micro-flares. Note that the second micro-flare is very tiny, barely above noise level on the GOES light curve. From the red time profiles in panels L1-L8, which stand for integrated intensity over the sunspot (red boxes in panels S1-S8) in 10830 \AA\ and AIA EUV wavelengths, we can see that there appear simultaneous EUV flashes and enhanced 10830 \AA\ absorption. Actually, the enhanced absorption at 21:31:00 UT has already appeared in Figure 4. The EUV flashes over the sunspot certainly have a spatio-temporal association with the two valleys on the 10830 \AA\ light curve, and at the same time, they are associated with the illumination of the two micro-flares.  Though we need to carry out further detailed quantitative analysis with existing models, the nice correlation will serve as a direct observational evidence supporting that PRM mechanism really works. Here it is worth mention that Judge and Pietarila (2004) reported no correlations between helium and He II emission and overlying coronal emission. Their conclusions were made by analyzing quiet regions and coronal holes. Considering in the paper that the two micro-flares and the excitation site are over 5 Mm away, the excitation of Helium atoms by EUV illumination should be a mean effect from overlying coronal emission as well as at least nearby coronal emission.

As mentioned in previous sections, the most notable phenomenon is that an oscillating absorption appears inside one flare ribbon when it sweeps into the sunspot.  An overall description of this phenomenon is that a dark (absorption) strip rushes out into the sweeping forth sunspot ribbon and then returns back until next several dark strips repeat the same behavior (Figure 1, also please watch the on-line movie). From the space-time diagram, the oscillating behavior exhibits itself as a sawtooth-like pattern, the oscillating absorption strip actually starts when the ribbon reaches the inner penumbral region. For umbral flashes and even penumbral waves, the sawtooth behavior can be observed from spectrograms taken in certain chromospheric lines like Ca\,\textsc{ii} H \& K lines (Beckers \& Tallant 1969). It was also called Z pattern with fast blueshift and slower redshift excursions, which were attributed to shock waves and post-shock falling back of the material contained in these shock waves. In this event, the observations were made in He\,\textsc{i} 10830 {\AA} blue wing -0.25 {\AA}, which certainly highlights the blushift component. However, the 0.5 {\AA} bandpass of the Lyot filter is comparable to the He\,\textsc{i} 10830 {\AA} line width ($\sim$ 0.75 {\AA}) and the line is very weak, slower redshift excursions will still be visible. So the features of blueshift and redshift excursions are showing up in the images of He\,\textsc{i} 10830 {\AA}.

We have to account for the material that falls back being visible only after the flare. Considering the magnitude (Figure 3) of the absorption, it is not the result of enhanced contrast against the bright background of flare ribbon. Enhanced absorption during the flare may play an important role for making slower redshift excursions come into visible absorption in the filtergrams.  On the other hand, the filament eruption will stretch the magnetic lines of force connected to the sunspot. Thus, the magnetic lines of force become more vertical and more material will fall back onto the sunspot.  This might be another possible reason why we can observe the falling behavior only after the flaring.

\section{SUMMARY} \label{s-sum}
By analyzing high resolution He\,\textsc{i} 10830 \AA\ narrow-band (NB) imaging data for a M 1.8 class two-ribbon flare on July 5, 2012 observed with the 1.6 meter aperture New Solar Telescope at the Big Bear Solar Observatory, we have made the following as initial findings:  1) Running penumbral waves and umbral flashes are manifest in absorption. From a space-time image made of a slit cutting across the sunspot's umbra and penumbra, running penumbral waves are directly seen as an outward extension of umbral flashes. To our knowledge, this is the first report of this phenomenon in the upper chromosphere from direct imaging. 2) The 10830 {\AA} absorption in the sunspot area gets enhanced during the flare. The enhancement is the result of the flare's EUV illumination. The result will serve as a direct observational evidence supporting the photo-ionization-recombination mechanism (PRM) for the excitation of Helium atoms. 3) The dark lanes for umbra flashes and penumbral waves in the space-time image obviously get broadened after the flare. We tentatively explained it as the result of a sudden increase in the density of ortho-Helium atoms in the sunspot area. However, we cannot exclude the possibility that the phenomenon reflects some intrinsic changes in the driving sources of umbral flashes during flaring. 4) A oscillating absorption strip appears inside one ribbon of the flare when it sweeps into the sunspot's penumbral and umbral regions. The oscillating absorption strip is the result of outward rushing umbral flashes and penumbra waves (both as shock waves) into the inwardly sweeping ribbon and the subsequent post-shock falling back of the material contained in these shock waves.

These kinds of phenomena was never observed in the low-resolution era, and they are surprising results of 10830 {\AA} high resolution NB imaging of solar flaring.
For the aforementioned phenomena, our explanations are only preliminary, further high-resolution observations of sunspot oscillations during flaring is needed. For example,
the oscillating absorption strip may be a useful tool to diagnose flare dynamics as well as sunspot dynamics and their possible relationships.  In this regard, we have succeeded in finding a plausible linkage between flare dynamics and sunspot dynamics. Simultaneous time series of several spectral lines, formed at different heights from the photosphere to the chromosphere, would be a powerful tool for studying the behavior of sunspot dynamics during flares.

\acknowledgements
The AIA data used here are courtesy of SDO (NASA) and the AIA consortium. We thank the AIA team for the easy access to calibrated data. We are very grateful to the referee for helping to improve the paper. This work was supported by NSFC grants 11333009, 10428309 and 11473071. Yingna Su was partly supported by the One Hundred Talent Program of Chinese Academy of Sciences. Philip R. Goode and Wenda Cao gratefully acknowledge support of AFOSR-FA 9550-15-1-0322. The BBSO operation is supported by NJIT, US NSF AGS 1250818, and NASA NNX13AG14G grants, and the NST operation is partly supported by the strategic priority research program of CAS with Grant No. XDB09000000 and by the Korea Astronomy and Space Science Institute and Seoul National University.

\newpage

\begin{figure}
\epsscale{.80}
\plotone{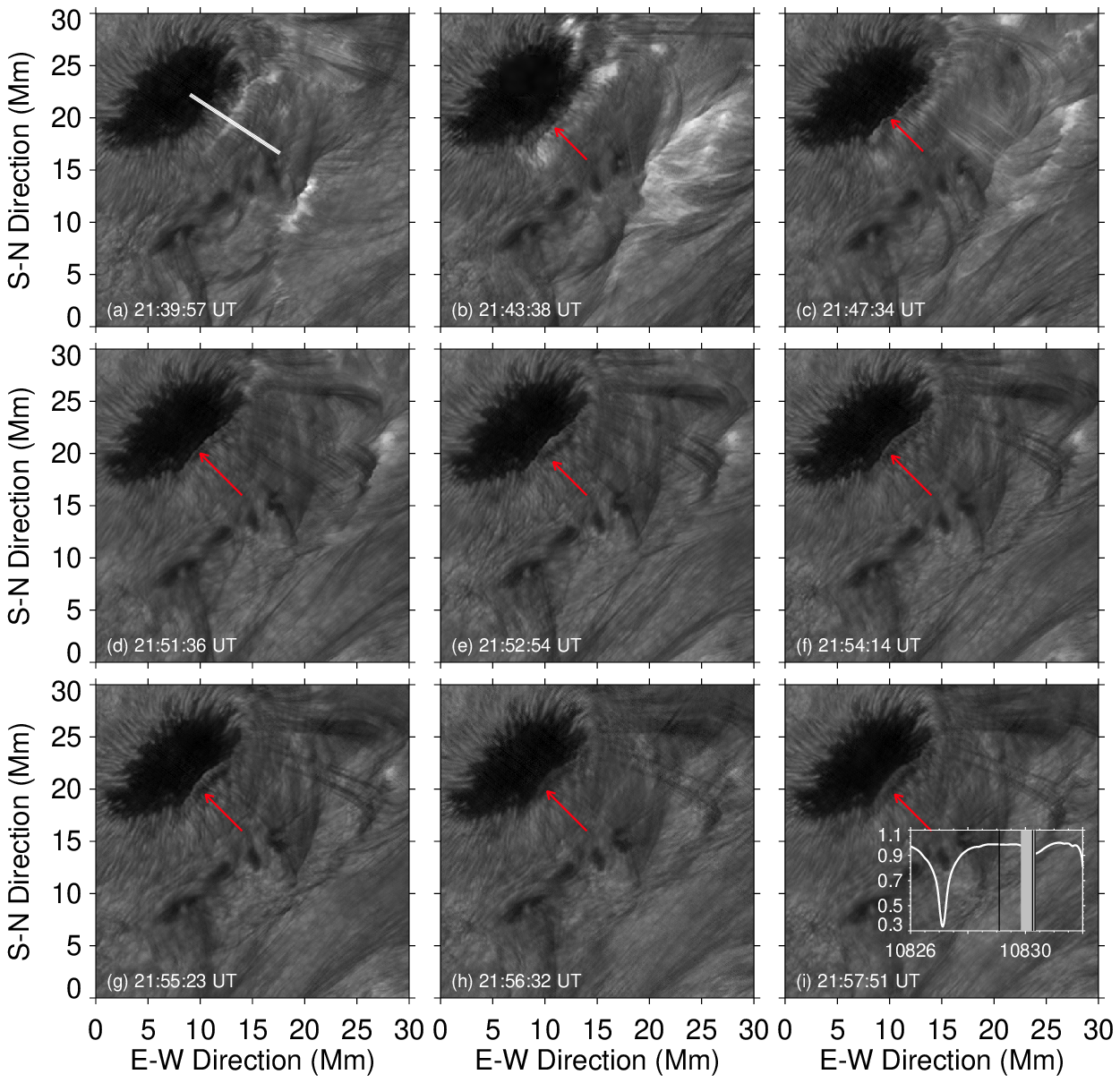}
\caption{BBSO/NST 10830 {\AA} narrow band filtergrams, showing the entire flaring process.  Note the swaying absorption strip inside the ribbon that rushed into the sunspot area, which is indicated by red arrows. They are speckle-reconstructed images. An animation of this figure is available in the online journal. The white line in panel (a) depicts the position of the slice for making the space-time image in Figure 2.
\label{fig1}}
\end{figure}

\clearpage

\begin{figure}
\epsscale{.80}
\plotone{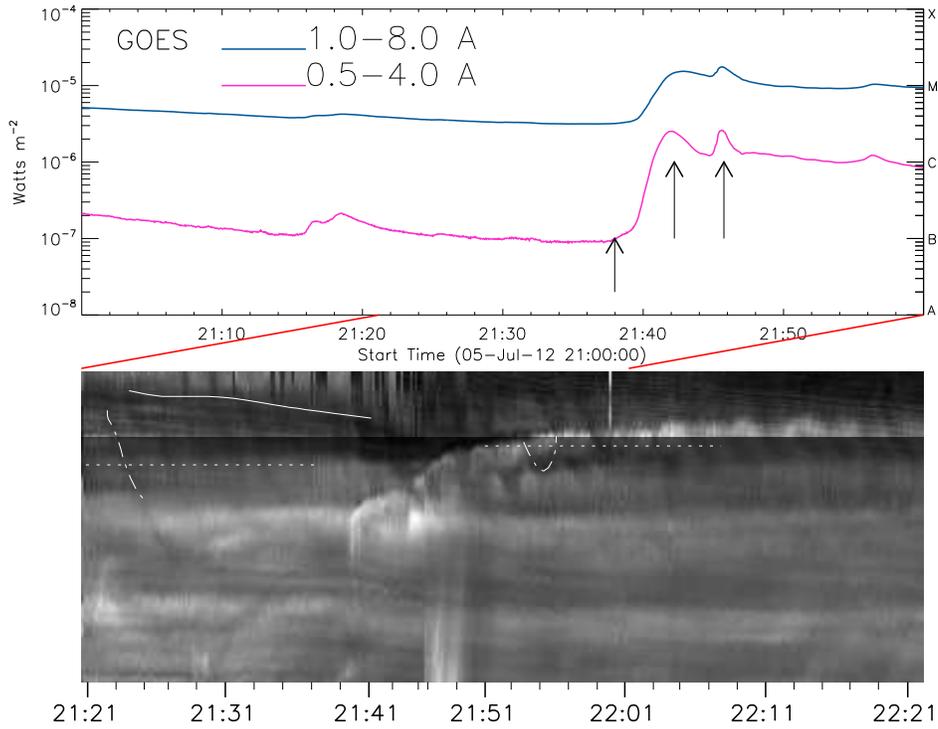}
\caption{Top panel: GOES X-ray light curve with the soft X-ray emission in the 1-8 \AA\ channel. The flare begins at about 21:38 UT and peaks at 21:42:15 UT and at 21:45:47 UT, which were each marked by arrows. Bottom panel: the space-time diagram, which was made from speckle-reconstructed images along a slice show in Figure 1a. The two white dash-dotted lines delineate the fishbone-like pattern and the sawtooth-like pattern. The two white dotted lines indicate the location of the spatial profile for wavelet analysis shown in Figure 3. The obvious interface indicates the boundary between sunspot's umbral and penumbral region, displayed with different gray scales. The irregular pattern above the solid line shows the area of umbral flashes distorted by speckle-reconstruction.
\label{fig2}}
\end{figure}

\clearpage

\begin{figure}
\epsscale{.80}
\plotone{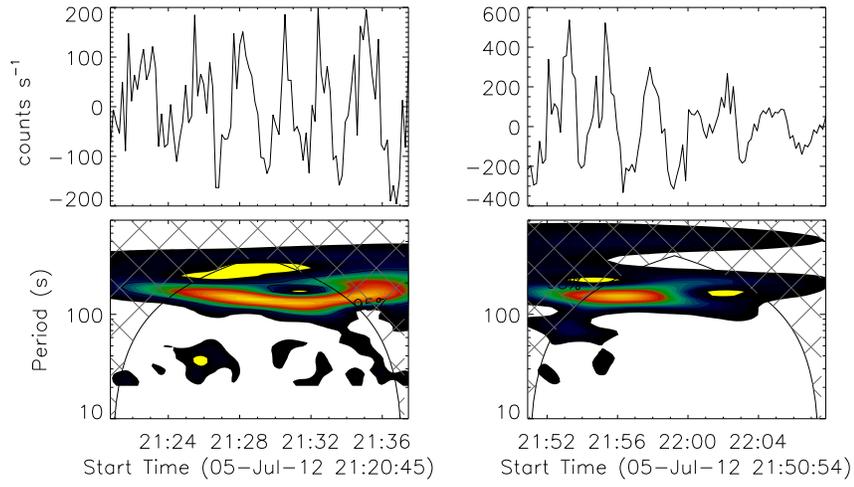}
\caption{Upper panels: the fast component for the intensity obtained from the space-time image along the two dotted lines (Figure 2). Lower panel shows their wavelet before and after the flare.
\label{fig3}}
\end{figure}

\clearpage

\begin{figure}
\epsscale{.80}
\plotone{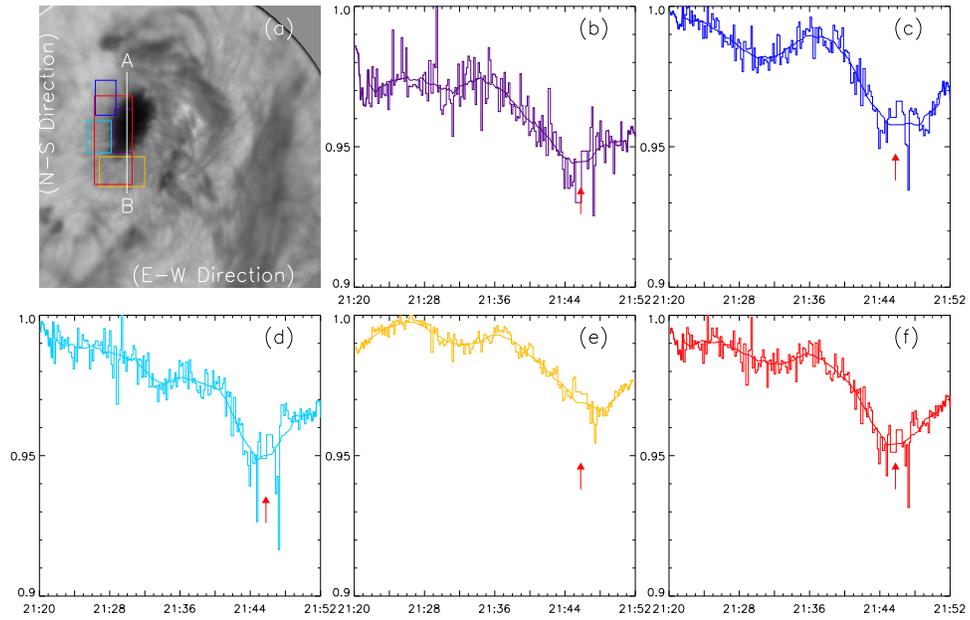}
\caption{The light curves of the five different regions in the sunspot area indicated with five color boxes in the first image. The light curves were obtained from lucky images without speckle reconstruction. The five regions include one umbral region, three different penumbral regions and one region covering the umbral and penumbral region. The three arrows in each panel mark the onset and two peaks of the flare.
\label{fig4}}
\end{figure}

\clearpage

\begin{figure}
\epsscale{.80}
\plotone{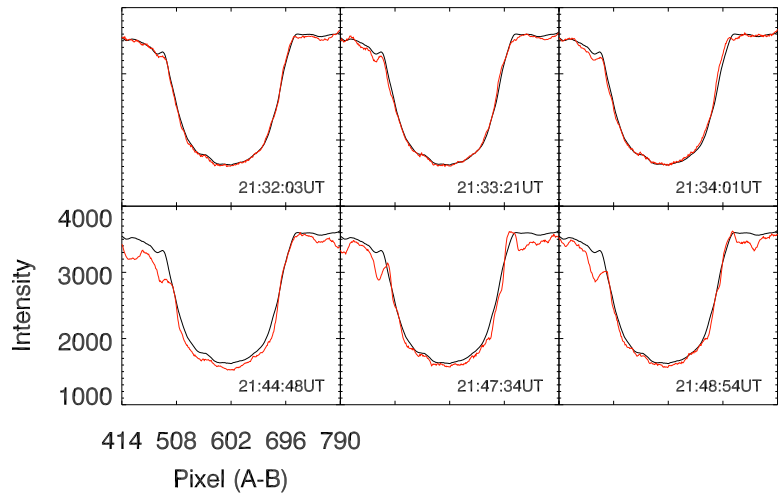}
\caption{The spatial profiles (red) for the intensity of the sunspot at certain times along the slice A-B depicted in Figure 4. The profile in black line in each panel is the average profile of 101 profiles from 21:20:06 UT to 21:36:58 UT before onset of the flaring.
\label{fig5}}
\end{figure}

\clearpage

\begin{figure}
\epsscale{.60}
\plotone{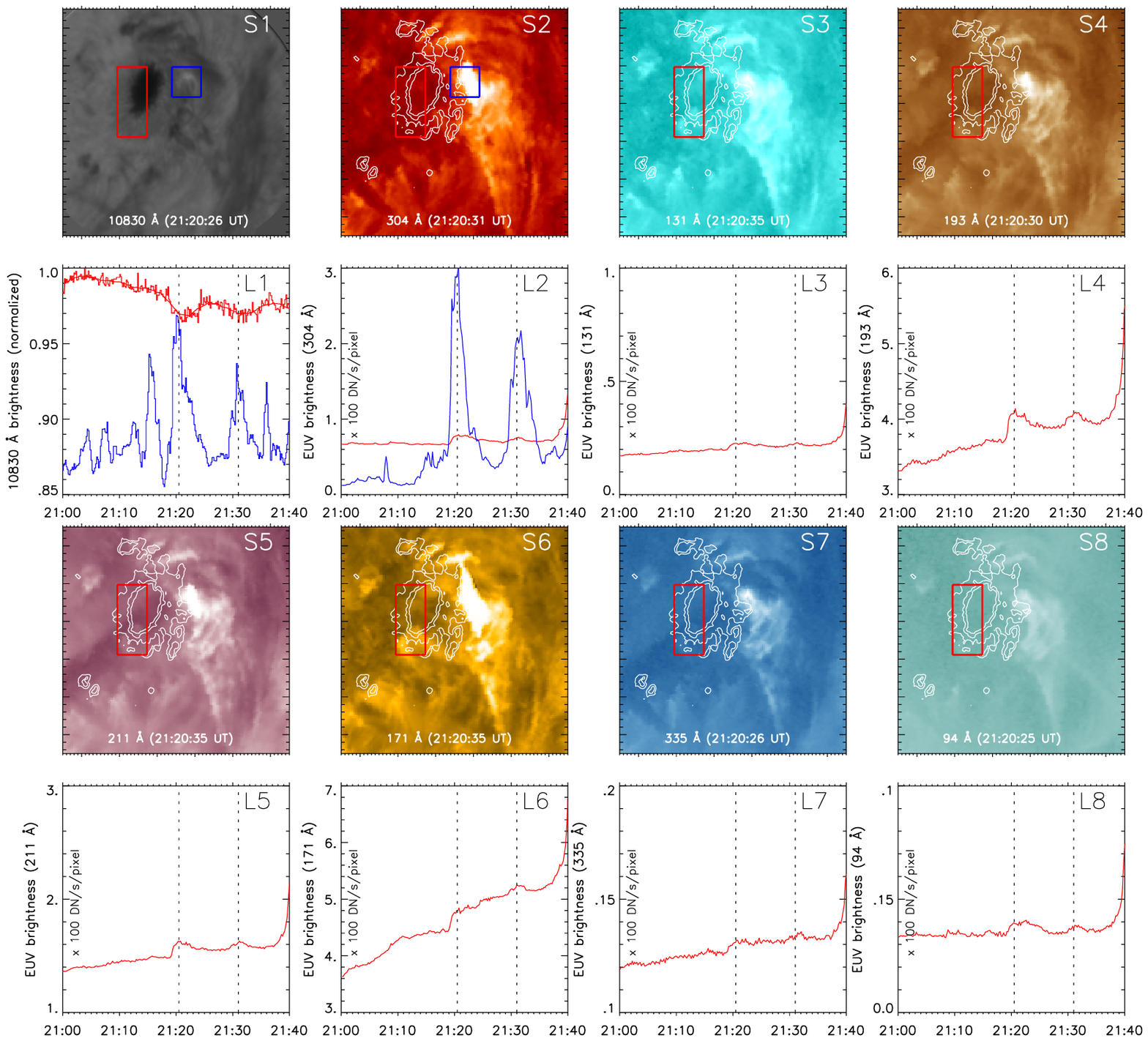}
\caption{Top and the 3rd rows (panels S1-S8): snapshots for the micro-flare at $\sim$ 21:20:30 UT taken at ground by NST and in space by AIA. AIA images are overlaid with contours of sunspots.  The 2nd and 4th rows  (panels L1-L8): light curves for 10830 \AA\ and EUV brightness from 21:00 UT to 21:40 UT obtained from the boxed areas shown in upper corresponding images, with red curves from red boxes (over the sunspot) and blue curves from blue ones (over the sites of the two micro-flares).  Two vertical dotted lines indicate the peak times of the two micro-flares.
\label{fig6}}
\end{figure}


\begin{thebibliography}{}
    \bibitem[]{}Andretta, V., $\&$ Jones, H. P. 1997, ApJ, 489, 375
    \bibitem[Avrett et al.(1994)]{1994isp..book...35A} Avrett, E.~H., Fontenla, J.~M., \& Loeser, R.\ 1994, Infrared Solar Physics, IAU Symp.~No.~154, eds.~D.M.~Rabin, J.T.~Jefferies, and C.~Lindsey, Kluwer, Dordrecht, pp.~35-47, 154, 35
    \bibitem[]{}Beckers, J. and Tallant, P. 1969, Solar Physics, 7., 351
    \bibitem[]{}Cao, W., Gorceix, N., Coulter, R., et al.\ 2010a, Astronomische Nachrichten, 331, 636
    \bibitem[]{}Cao, W., Gorceix, N., Coulter, R., Coulter, A., \& Goode, P.~R.\ 2010b, \procspie, 7733, 773330
    \bibitem[Elmore et al.(1998)]{1998ApOpt..37.4270E} Elmore, D.~F., Card, G.~L., Chambellan, C.~W., et al.\ 1998, \ao, 37, 4270
    \bibitem[Fang et al.(1998)]{2013RAA....13.1509}Fang, C. et al. 2013, RAA, 13, 1509
    \bibitem[Felipe et al.(2010)]{2010ApJ...722..131F} Felipe, T., Khomenko, E., Collados, M., \& Beck, C.\ 2010, \apj, 722, 131
    \bibitem[]{}Goode, P. R., Coulter, R., Gorceix, N., Yurchyshyn, V., \& Cao, W. 2010, Astron. Nachr., 331, 620
    \bibitem[Harvey et al.(1996)]{1996AAS...188.6703H} Harvey, J., Keller, C., November, L., \& NSO Staff 1996, Bulletin of the American Astronomical Society, 28, 67.03
    \bibitem[]{}Ji, H., Cao, W., \& Goode, P.~R.\ 2012, \apjl, 750, L25
    \bibitem[]{}Ji, H., You, J., and Fan, Z. 1997, ChA\&A, 21, 347
   \bibitem[Judge et al.(2015)]{2015ApJ...814..100J} Judge, P.~G., Kleint, L., \& Sainz Dalda, A.\ 2015, \apj, 814, 100
   \bibitem[Judge \& Pietarila(2004)]{2004ApJ...606.1258J} Judge, P.~G., \& Pietarila, A.\ 2004, \apj, 606, 1258
   \bibitem[Lemen et al.(2012)]{2012SoPh..275...17L} Lemen, J.~R., Title, A.~M., Akin, D.~J., et al.\ 2012, \solphys, 275, 17
    \bibitem[Lites(1986)]{1986ApJ...301.1005L} Lites, B.~W.\ 1986, \apj, 301, 1005
    \bibitem[Neidig et al.(1998)]{1998ASPC..140..519N} Neidig, D., Wiborg, P., Confer, M., et al.\ 1998, Synoptic Solar Physics, 140, 519
    \bibitem[]{}Penn,M.J. and Jones,H.P., 1996, Solar Physics, 168, 19
    \bibitem[]{}Pesnell, W.~D., Thompson, B.~J., \& Chamberlin, P.~C.\ 2012, \solphys, 275, 3
    \bibitem[]{}Rouppe van der Voort, L.~H.~M., Rutten, R.~J., S{\"u}tterlin, P., et al.\ 2003, \aap, 403, 277
    \bibitem[]{}Tian, H., DeLuca, E., and Reeves, K.K. et al. 2014, ApJ, 786, 137
    \bibitem[]{}Thomas, J,H., Cram, L.E., and Nye, A.H. 1984, ApJ, 285, 368
    \bibitem[Woods et al.(2012)]{2012AGUFMSH41A2099W} Woods, T.~N., Woodraska, D., Jones, A.~R., Eparvier, F.~G., \& Caspi, A.\ 2012, AGU Fall Meeting Abstracts
    \bibitem[]{}W{\"o}ger, F., \& von der L{\"u}he, O.\ 2007, \ao, 46, 8015
    \bibitem[Yurchyshyn et al.(2015)]{2015ApJ...798..136Y} Yurchyshyn, V., Abramenko, V., \& Kilcik, A.\ 2015, \apj, 798, 136
    \bibitem[Zeng et al.(2014)]{2014ApJ...793...87Z} Zeng, Z., Qiu, J., Cao, W., \& Judge, P.~G.\ 2014, \apj, 793, 87
    \bibitem[Zeng et al.(2013)]{2013ApJ...769L..33Z} Zeng, Z., Cao, W., \& Ji, H.\ 2013, \apjl, 769, L33
\end{thebibliography}
\end{document}